\documentclass[aps,pra,twocolumn,superscriptaddress]{revtex4}
\usepackage{graphicx}
\usepackage{amsmath}
\usepackage{amsfonts}%
\usepackage{amssymb}
\usepackage{mathrsfs}
\usepackage{dcolumn}% Align table columns on decimal point
\usepackage{bm}% bold math
\usepackage{float}
\usepackage{color}
\usepackage{epsfig}% FOR CORRECT FIGURE CAPTIONS

\begin{document}
\title{Supercontinuum  generation with bright and dark solitons in optical fibers}
\author{C. Mili\'{a}n}
%\thanks{These two authors contributed equally to this work}
\email{carles.milian@cpht.polytechnique.fr; tomy.marest@ed.univ-lille1.fr}
\affiliation{Centre de Physique Th\'{e}orique, CNRS, \'{E}cole Polytechnique, F-91128 Palaiseau, France}
\author{T. Marest}
\email{carles.milian@cpht.polytechnique.fr; tomy.marest@ed.univ-lille1.fr}
%\thanks{These two authors contributed equally to this work}
\affiliation{Universit\'e Lille, CNRS, UMR 8523--PhLAM--Physique des Lasers Atomes et Mol\'ecules, F-59000 Lille, France}
\author{A. Kudlinski}
\affiliation{Universit\'e Lille, CNRS, UMR 8523--PhLAM--Physique des Lasers Atomes et Mol\'ecules, F-59000 Lille, France}
\author{D. V. Skryabin}
\affiliation{Department of Physics, University of Bath, Bath BA2 7AY, United Kingdom}
\affiliation{ITMO University, Kronverksky Avenue 49, St. Petersburg 197101, Russian Federation}
\begin{abstract}
We study numerically and experimentally supercontinuum generation in optical fibers with dark and bright solitons simultaneously contributing into the spectral broadening and dispersive wave generation.  We report a novel type of weak trapped radiation arising due to interaction of bright solitons with the dark soliton background. This radiation expresses itself as two  pulses  with the  continuously shifting spectra constituting  the short and long wavelength limits  of the continuum. Our theoretical and experimental results are in good agreement.
\end{abstract}
\maketitle
%

% INTRO & PAPER
Supercontinuum  generation in optical fibers has mostly been associated with bright solitons and their dispersive radiation \cite{DudleyRMP,SkryabinRMP}. When  spectra are at their maximum span, they typically consist of a number of bright solitons with different carrier frequencies in the spectral range with anomalous group velocity dispersion (GVD) and various types of dispersive waves (DWs) in the normal or sometimes anomalous GVD ranges, for example Cherenkov \cite{AkhmedievPRA,BiancalanaPRE} and Airy \cite{GorbachOE} resonant radiation, emitted by these solitons. On the other hand, the role of dark solitons in fiber supercontinuum remains relatively unexplored, despite the substantial knowledge about isolated dark solitons and their dynamics under the typical perturbations present in the nonlinear propagation of light in fibers \cite{KivsharPR,OreshnikovOL}. Recently we have demonstrated experimentally \cite{MarestOL} and previously numerically \cite{MilianOL}, that a conceptually analogous supercontinuum generation picture to that involving bright solitons is achieved when formation of dark solitons rules the spectral broadening.

In this work, we present for the first time supercontinuum generation in photonic crystal fibers (PCFs) combining both bright and dark solitons. We achieve coexistence of the two types of solitary waves by pumping the fiber in the normal GVD range with two delayed sub-picosecond pulses giving rise to a train of dark solitons when higher order linear and nonlinear effects are negligible during the first stages of the propagation \cite{RothenbergOC,RothenbergOL,FinotOFT}. After some propagation length, the red-detuned tail of the dark soliton train starts overlapping with the anomalous GVD region. For the high input powers, this tail undergoes its own nonlinear dynamics and develops a shock front which generates several bright solitons \cite{Bose2015} (see also numerical results in Refs. \cite{ConfortiOL,ConfortiPRA}). The spectral content of the supercontinuum is further enriched by the emission of dispersive Cherenkov radiation by  bright and dark solitons.

In addition to the above dynamics, we report in this work a particular type of trapped radiation that appears due to the interaction of the bright solitons with the broad intense pulse hosting a train of dark solitons in the normal GVD range. This radiation has two spectral components and manifests itself through continuous red and blue shifts. Features of this radiation are its growth in the absence of phase matching and its subsequent evolution without the velocity matching condition with the carrier of the bright soliton under which trapping is commonly studied \cite{SkryabinRMP}.
% FIG 1
\begin{figure*}
\includegraphics[width=.9\textwidth]{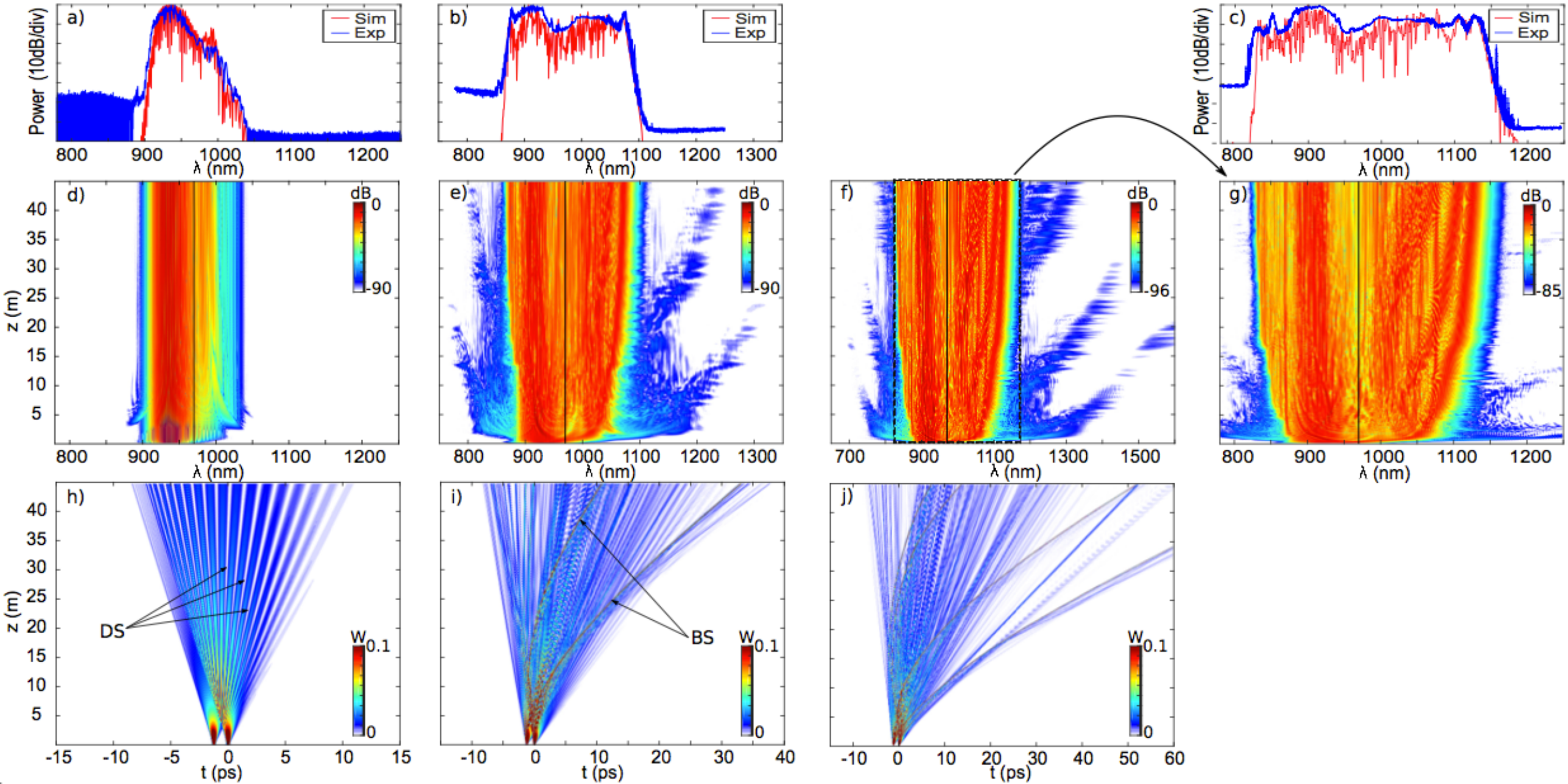}
\caption{Bright-Dark soliton supercontinuum generation. (a-c) Comparison between numerical (red) and experimental (blue) results at the fiber output, $z=45$ m. (d-g) Spectral and (h-j) temporal evolution in a PCF with pitch $\Lambda=2.78\ \mu$m and air filling fraction $d/\Lambda=0.593$. (g) is a zoom of the central part of (f). Input power from left to right: $P_0= $150 W, 860 W, and 1600 W. Other input conditions are: pump wavelength $\lambda_0=936$ nm, input pulse delay $t_{del}=1.25$ ps, and pulse duration $T_0=137$ fs. Vertical lines in spectral plots mark the zero GVD wavelength, $\lambda_{zgvd}=970$ nm. Labels DS (BS) stand for dark (bright) solitons. Experiments were performed with the same setup as in \cite{MarestOL}.\label{f1}}
\end{figure*}
We assume that the propagation of the electric field envelope, $A$, is governed by the nonlinear Schr$\ddot{\mathrm{o}}$dinger (NLS) equation augmented with the experimental waveguide dispersion and Raman effect,
\begin{eqnarray}
&& -i\partial_Z A=-\mathcal{\hat D}_{\omega_0}A+(1-f_R)|A|^2A+QA \label{eq:fiber}, \\ &&
\mathcal{\hat D}_{\omega_0}\equiv L_D\sum_{m\geq1}\frac{\beta_m}{m!}(-i\partial_t)^m, \\&&
\partial_t^2Q+2\Gamma_R\partial_tQ+\left[\omega_R^2+\Gamma_R^2\right]\left[Q-f_R|A|^2\right]=0, \label{eq:raman}
\end{eqnarray}
where $\mathcal{\hat D}_{\omega_0}$ is the  dispersion operator and $\beta_m\equiv\partial^m\beta(\omega)/\partial\omega^m\vert_{\omega_0}$ are the Taylor expansion coefficients of the propagation constant, $\beta(\omega)$, of the fiber modes evaluated at the input frequency $\omega_0\equiv2\pi c/\lambda_0$, $T_0$ is the initial pulse width, and $L_D\equiv T_0^2/|\beta_2|$ is the dispersion length. $Z$ is the propagation distance measured in units of $L_D$: $Z=z/L_D$, being $z$ the physical length. $Q$ is the Raman part of the nonlinear susceptibility \cite{ShenPR65,BoydBOOK} with the standard parameters used for silica glass: $f_R=0.18$, $\Gamma_R=T_0(fs)/32$, $\omega_R=T_0(fs)/12.2$. Since the field $A$ in the fiber satisfies $\lim_{t\rightarrow-\infty}A=0$, $Q$ satisfies $\lim_{t\rightarrow-\infty}\{Q,\partial_tQ\}=\{0,0\}$ and Eq.\ref{eq:raman} can be solved analytically:
\begin{eqnarray}
Q=f_R\frac{\omega_R^2+\Gamma_R^2}{\omega_R}|A|^2\ast\Big(\exp{\left(-\gamma_Rt\right)}\sin\left({\omega_Rt}\right)\Theta(t)\Big),
\end{eqnarray}
where '$\ast$' denotes convolution product and $\Theta$ is the Heaviside function. 
% FIG 2
\begin{figure}
\includegraphics[width=.5\textwidth]{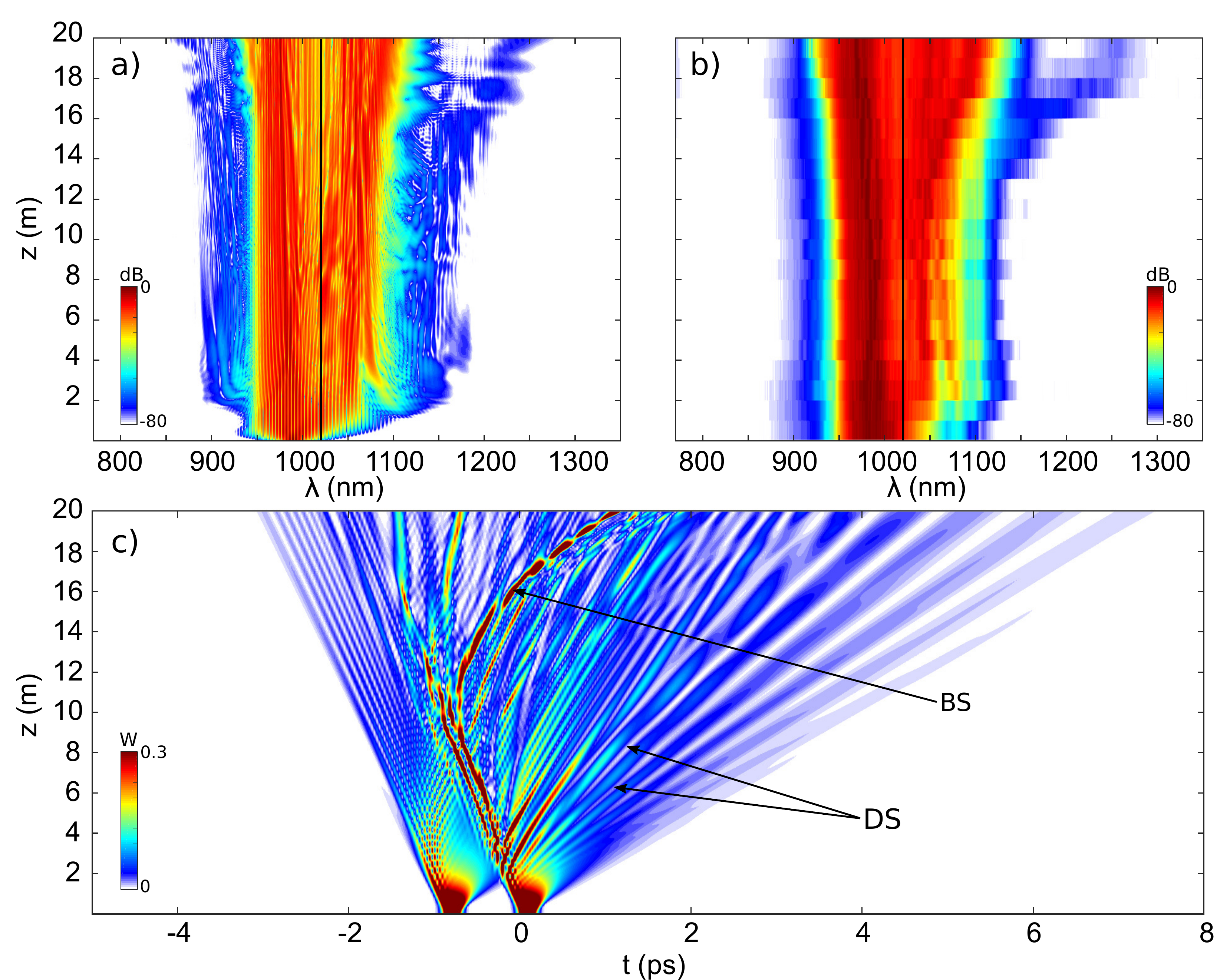}
\caption{(a,c) Spectral and temporal numerical evolution over 20 m. (b) Experimentally recorded spectral evolution by the cutback technique. Experiments were carried out on the setup described in Ref. \cite{MarestOL} in a PCF with a zero dispersion wavelength of $\lambda_{zgvd}=1021$ nm. Input parameters: $\lambda_0=989$ nm, $P_0=685$ W, $\tau_0=137$ fs, and $t_{del}=880$ fs.\label{f2}}
\end{figure}

Figure \ref{f1} shows supercontinuum generation from two delayed Gaussian pulses in the form
\begin{eqnarray}
&& A(t,z=0)=\\ && \nonumber
\sqrt{P_0}\left[\exp\left\{-\left(\frac{t+t_{del}}{\sqrt{2}T_0}\right)^2\right\}+\exp\left\{-\left(\frac{t}{\sqrt{2}T_0}\right)^2\right\}\right],\label{eq:Ain}
\end{eqnarray}
where $t_{del}$ is the temporal delay, $T_0=\tau_0/1.665$, and $\tau_0$ is the full width at half maximum (FWHM) of the transform limited pulses.
Figures \ref{f1}(a,d,h) show that the spectral broadening for the relatively low input powers is mainly ruled by the formation of dark solitons and their dynamics under perturbations \cite{KivsharPR}. The most notable effect is the significant transfer of energy to the anomalous GVD range by means of DWs. The strongest DWs come from the dark solitons having large amplitude background and the carrier frequency close to the zero GVD. The latter condition is satisfied for the fastest dark solitons \cite{MilianOL,MarestOL}. In Figs. \ref{f1}(a,d,h) the dark soliton train remains mostly in the normal GVD  and therefore bright solitons cannot be formed. However, for the higher input power, this pictures changes qualitatively, as can be seen in Figs. \ref{f1}(b,c,e-g,i,j). The stronger overlap of the red-shifted tail of the dark soliton train with the anomalous GVD yields the formation of several fundamental bright solitons, nesting on the fast edge of the dark soliton train, whose temporal and spectral signatures are evident in the propagation plots. Considering that the number of dark solitons tends to increase with the input power \cite{FinotOFT}, for moderate and high power levels, around $P_0\gtrsim400$ W in our PCF, a mixed dark and bright soliton dynamics takes place yielding a rich spectral content at both sides of the zero GVD wavelength.

An important feature of supercontinuum that we report here, is that once formed, the bright solitons then go through the whole dark soliton train as they are slowed down by Raman and recoil effects. As will be shown later, this interaction of bright and dark solitons results in additional spectral components associated to DWs emitted from the bright solitons, and other weak waves that remain trapped by the bright solitons and quickly shift towards significantly smaller and greater wavelengths than the rest of the spectrum. In particular, the red shifted part of this radiation was recorded in experiments, see Fig. \ref{f2}.

\begin{figure}
\includegraphics[width=.5\textwidth]{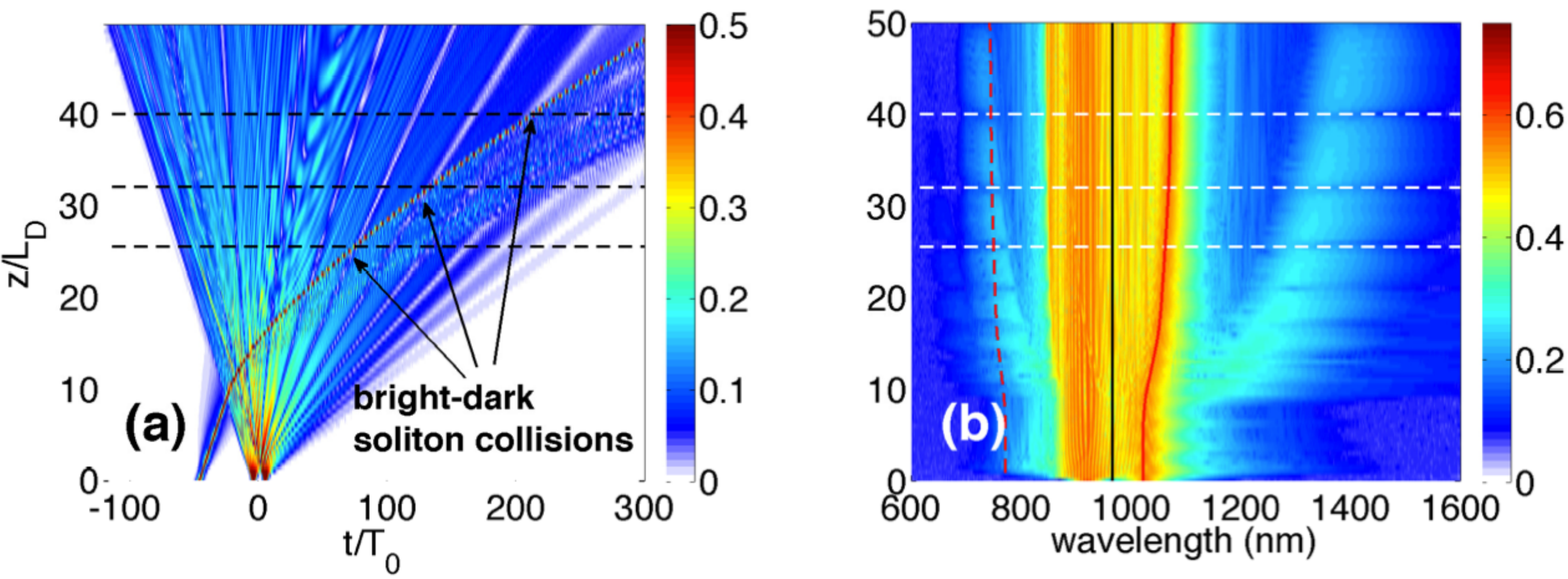}
\caption{(a) temporal and (b) spectral evolution of a bright soliton colliding with a dark soliton train. Input conditions for the dark soliton train consist on two delayed Gaussian pulses with $T_0=50$ fs, $t_{del}=465$ fs, $P_0=1.97$ kW at $\lambda_0=920$ nm. At this wavelength %$\beta_0=9.85\times10^9$ 1/km, $\beta_1=4.9\times10^6$ ps/km,
$\beta_2=5.89$ ps$^2$/km and $\beta_3=6\times10^{-2}$ ps$^3$/km ($\lambda_{zgvd}=964.8$ nm). The bright soliton is launched at $t\approx-50T_0$, $\lambda_s=925$ nm, and $P_s=1.13$ kW ($N\approx1$). Dashed horizontal lines mark the bright-dark soliton collisions in time and frequency domains. Red curves in (b) mark the soliton carrier (full) and the wacelength of the velocity matched waves (dashed).
\label{f3}}
\end{figure}

In order to explain the above mentioned aspects of the dynamics in more details, we show numerical simulations in which an initially isolated bright soliton collides and propagates through the dark soliton train which peak power is sufficiently low so it does not give birth to additional bright solitons [c.f. Fig. \ref{f3}]. To gain  further insight  we  expand the field $A$ as:
\begin{equation}
A=\Psi_S+\Psi_b+\Psi_d+g,
\end{equation}
where subscripts $S,b,d$ refer, respectively, to the bright soliton, the nonlinear background containing all dark solitons, and the dispersive waves emitted by the dark solitons early in the dynamics. Here, $g$ accounts for the energy transfer between $\Psi_{S,b,d}$.  By substituting the above \textit{anstaz} in Eq. \ref{eq:fiber} with $f_R=0$, we write the propagation equations for the four different field components:
\begin{eqnarray}
 -i\partial_z\Psi_S&+&\mathcal{\hat{D}}_{\omega_S}^{(2)}\Psi_S-[|\Psi_S|^2+2(|\Psi_b|^2+|\Psi_d|^2)]\Psi_S=0\label{eq:psiS},\\
-i\partial_z\Psi_b&+&\mathcal{\hat{D}}_{\omega_b}^{(2)}\Psi_b-[|\Psi_b|^2+2(|\Psi_S|^2+|\Psi_d|^2)]\Psi_b=0\label{eq:psib},\\ \nonumber
-i\partial_z\Psi_d&+&\mathcal{\hat{D}}_{\omega_d}\Psi_d-[|\Psi_d|^2+2(|\Psi_b|^2+|\Psi_S|^2)]\Psi_d=\mathcal{S}_{DW,b},\\ &&\label{eq:psid}\\
-i\partial_zg&+&\mathcal{\hat{D}}_{\omega_g}g-2|\Psi|^2g-\Psi^2g^*=\mathcal{S}_I+\mathcal{S}_{DW,S}\label{eq:g},\\
\mathcal{S}_I&=&\Psi_S^2\Psi_b^*+\Psi_b^2\Psi_S^*+\\ && \nonumber\Psi_d^*[\Psi_S+\Psi_b]^2+4\Psi_d\mathrm{Re}\{\Psi_b\Psi_S^*\}
+\Psi_d^2[\Psi_S^*+\Psi_b^*]\label{eq:SI},\\
\mathcal{S}_{DW,x}&=&[\mathcal{\hat{D}}_{\omega_x}-\mathcal{\hat{D}}_{\omega_x}^{(2)}]\Psi_x\label{eq:SRR},
\end{eqnarray}
where $\mathcal{D}_\omega^{(2)}$ denotes GVD of the wave at frequency $\omega$. 

In the first stages of the propagation, while the bright soliton remains isolated from the dark solitons ($z\lesssim10\ L_D$ in Fig. \ref{f3}), each dark soliton in the train, located at $t_{DS}$, is locally described by $\Psi_b(t\sim t_{DS})$ and emits DWs, $\Psi_d$, which are driven by ${\mathcal{S}_{DW,b}}\sim\partial_t^3\Psi_b(t\sim t_{DS})$ [see Eq. \ref{eq:psid}]. Frequencies of these dispersive waves are predicted by the resonance condition derived in Refs. \cite{KarpmanPLA,AfanasjevOL}, which have been proven to accurately match the experimental observations \cite{MarestOL}.
%
%FIG 3
%FIG 4
\begin{figure}
\includegraphics[width=.5\textwidth]{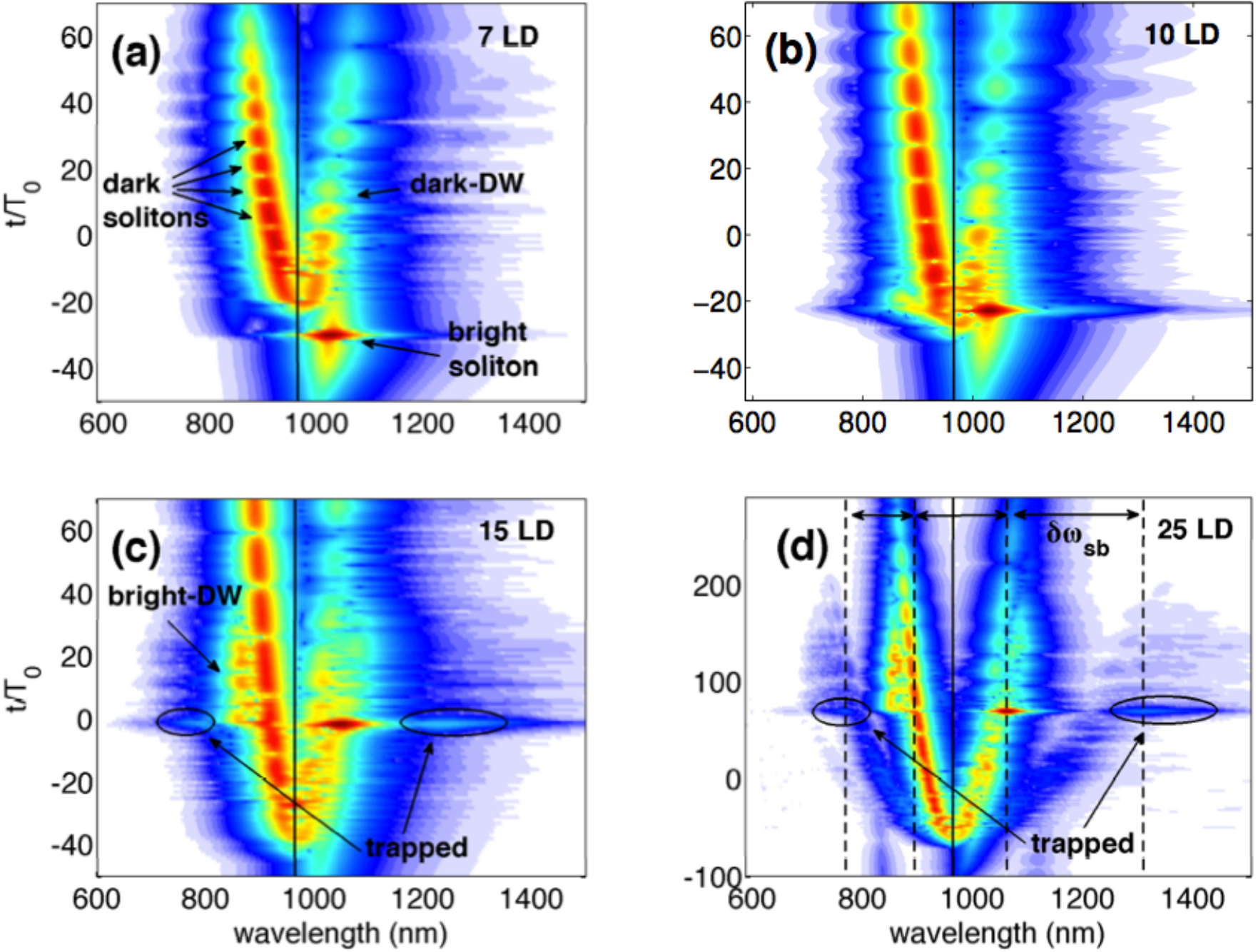}
\caption{XFROG diagrams corresponding to (a) z/$L_D$=7, (b) 10, (c) 15, and (d) 25 in Fig.\ref{f3}. Vertical solid lines represent the zero of the GVD. Vertical lines in (d) mark the instantaneous bright soliton wavelength and that of the background overlapping in time with it. Ellipses enclose the trapped pulses. \label{f4}}
\end{figure}
%FIG 5
\begin{figure}
\includegraphics[width=.5\textwidth]{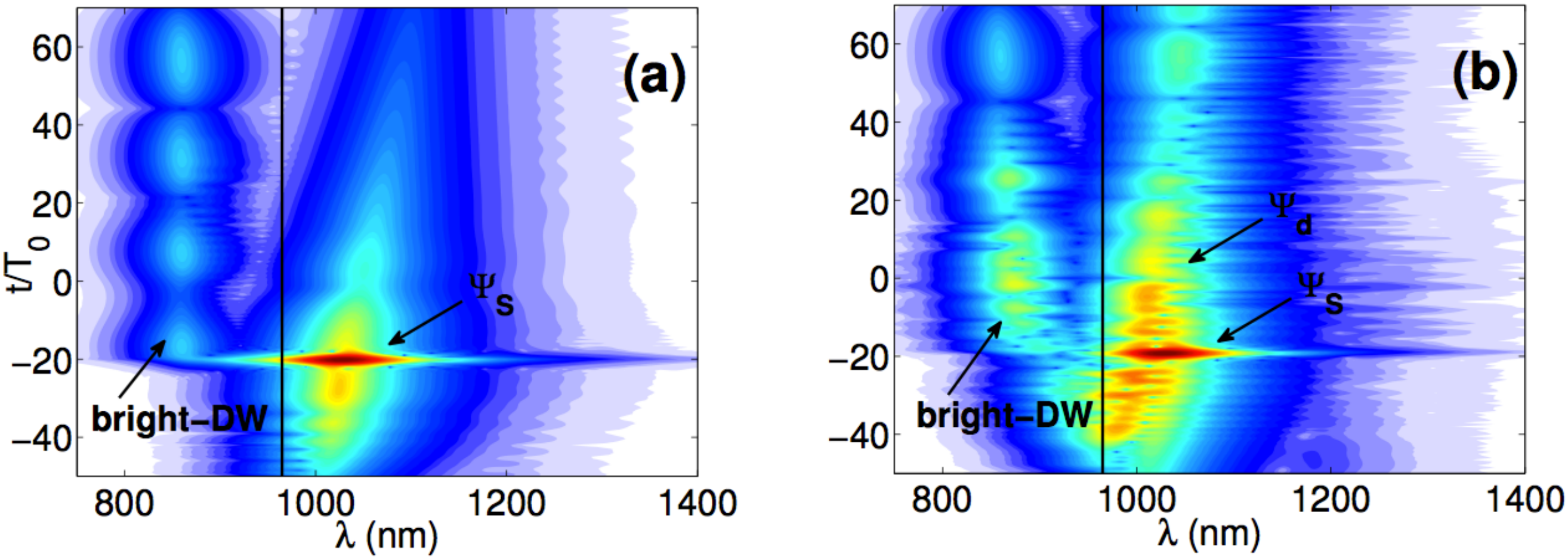}
\caption{XFROG diagrams corresponding to z=15 $L_D$ in Fig. \ref{f4}(c). Initial conditions are given by those at $z=7\ L_D$ in Fig. \ref{f4}(a) and filtering: (a) $\Psi_b$ and $\Psi_d$, (b) $\Psi_b$. Vertical solid lines represent the zero of the GVD. In (a), the DW is emitted in pulses because the $\Psi_S$ has order $1<N\lesssim2$ and it is therefore oscillatory (similarly to results in Refs. \cite{MilianOE14,MilianPRA})\label{f5}}
\end{figure}

At $z\approx10\ L_D$ the bright soliton collides with the dark soliton train, $\Psi_{b}$, and its DWs, $\Psi_d$ [see Fig. \ref{f3}(a)]. This collision produces relatively strong radiation around 900 nm, with wavelength shorter than the one of the dark soliton background. As apparent from XFROGs in Fig. \ref{f4} this radiation, labeled as \textit{bright-DW}, is emitted at the bright soliton temporal location and travels slower than it on top of the chirped background, $\Psi_b$. Also, note that the bright soliton spectrally recoils \cite{AkhmedievPRA} when this radiation appears (we recall simulations in Fig. \ref{f3} are done with $f_R=0$) and therefore this radiation is a dispersive wave emitted by the bright soliton. Because the bright soliton is too far detuned from the zero GVD to emit DWs efficiently, the origin of this DW is in the interaction with $\Psi_b$ and $\Psi_d$. Below, we provide further insights on this interaction.

Fig. \ref{f5} shows XFROG's at $z=15L_D$ when two filters have been applied at $z=7L_D$ to the field in Fig. \ref{f4}(a) (i.e., before the collision at $z\approx10\ L_D$ occurs). Figure \ref{f5}(a) shows the result of propagating $\Psi_S$ alone (filtering $\Psi_{b,d}$), and Fig. \ref{f5}(b) the result of $\Psi_S$ interacting with $\Psi_d$ (filtering only $\Psi_{b}$).
Comparison of Figs. \ref{f5}(a) and \ref{f5}(b) shows that the bright soliton emits DWs more efficiently when the process is fueled by the dark soliton DWs, $\Psi_d$, as it was studied in Ref. \cite{YulinOL}. This indicates that the source term for $g$ in Eq. \ref{eq:g} is $\mathcal{S}_{I}\approx\Psi_S^2\Psi_d^*$ at the first order. Furthermore, comparison of Fig.\ref{f5}(b) and Fig.\ref{f4}(c) shows that the presence of the nonlinear background, $\Psi_b$, further amplifies the DW's.
The above observations brings us to conclusion that the DW emitted by the bright soliton is fueled by the DW's emitted from the dark solitons, $\Psi_d$, and amplified by the nonlinear background that hosts the dark solitons, $\Psi_b$.

We now discuss the spectral edges of the continuum. From Figs. \ref{f3} and \ref{f4} it is clear that the collision of the bright soliton with the radiating train, and in particular with $\Psi_b$, results in the formation of the fast shifting trapped radiation. To understand this effect, we now focus our attention on the interaction terms in $\mathcal{S}_I$. Figure \ref{f4} shows that the dispersive waves emitted by the dark solitons are less powerful than the bright soliton and the background, i.e., $|\Psi_d|<|\Psi_S|,|\Psi_b|$, therefore suggesting the source term for $g$ in Eq. \ref{eq:g} may be approximated by the first two terms only: $\mathcal{S}_I\approx\Psi_S^2\Psi_b^*+\Psi_b^2\Psi_S^*$. This immediately reveals that the growth of $g$ occurs predominantly for the two frequencies $\omega_{t1}=2\omega_b-\omega_S$, $\omega_{t2}=2\omega_S-\omega_b$ [subindex $t$ stands for \textit{trapped}]. These two frequencies are offset by $\delta\omega_{sb}$ from $\Psi_b$ and by $-\delta\omega_{sb}$ from $\Psi_S$, where $\delta\omega_{sb}\equiv\omega_b-\omega_S$ [see Fig. \ref{f4}(d)], and correspond to the weak trapped radiations, $g_t$, observed in our modeling and experiments. Therefore, the two spectral branches $g_t$ exist solely because of the interaction of the bright soliton, $\Psi_S$, and the background, $\Psi_b$.

Along propagation, the soliton, $\Psi_S$, moves on top of the chirped background, $\Psi_b$, and it also recoils constantly due to the continuous DW emission. These two facts yield a changing frequency detuning $\delta\omega_{sb}$ along $z$. The weak radiations $g_t$ that is traveling exactly at the center (in time) of the soliton is then constantly \textit{invested} in generating the new $g_t$ at the right frequency. It is this fact that gives an apparent trapped character to $g_t$. This effect is a form of trapping, but not in the sense that all three waves $\Psi_S$, $g_t(\omega_{t1})$, and $g_t(\omega_{t2})$ have the same velocity. In fact, the velocities of the weak waves $g_t$ are far from being matched with that of the soliton $\Psi_S$, specially the red shifted component [see Fig.\ref{f3}(b)]. Consistently with the above, we observe that the existence of $g_t$ is strongly affected by the fact that $\Psi_b$ is decreasing: $\Psi_S$ overlaps with parts of the pulse $\Psi_b$ that decrease in power so when $\Psi_b$ gets smaller also does $g_t$. Indeed we observe that when the soliton $\Psi_S$ decouples from $\Psi_b$, $g_t$ decays to zero. As a consequence, during each bright-dark soliton collision $g_t$ faints substantially [see Fig. \ref{f3}].

From Eq. \ref{eq:g} it is clear that many other waves could have grown due to frequency mixing processes. We note the expansion of the source terms $\mathcal{S}$ (Eq. \ref{eq:SI}) in powers of $\Psi_d$ ($|\Psi_d|<|\Psi_S|,|\Psi_b|$) reveals that the terms of the order $\sim\Psi_d,\Psi_d^2$ are much less efficient and here we observe only growth represented by terms of Eq. \ref{eq:SI} which are at zeroth order in $\Psi_d$.

The inclusion of Raman scattering in our modeling, $f_R=0.18$ in Eq.\ref{eq:fiber}, is mainly seen to amplify the DWs emitted by the dark solitons \cite{MilianOL} and to enhance the shift $d \delta\omega_{sb}/dz$ of the trapped waves towards longer and shorter wavelengths, through the bright soliton induced frequency shift \cite{MitschkeOL,GordonOL}. Despite the little qualitative impact of Raman scattering in our results, its quantitative impact was indeed important to get a good agreement in between the experiments and simulations.

In summary, we have reported on supercontinuum generation by dark and bright solitons in optical fibers. Both types of solitons contribute to the spectral broadening via the emission of dispersive waves, collisions, and Raman induced frequency shift. We have shown that by varying the input power of the delayed pulses one can control the number of dark and bright solitons that are formed [see Fig.\ref{f1}]. Additionally, mixing of bright solitons with background of the dark soliton train results in two spectrally far detuned pulses that remain trapped in the bright solitons and define the low and high frequency edges of the supercontinuum. These results are important for supercontinuum generation from fs laser pulses that spectrally fall in the normal GVD of the waveguide.

{\bf{Funding.}} Direction G\'en\'erale de l'Armement (DGA);
Russian Foundation for Basic Research (RFBR); ITMO University. T. M. and A. K. acknowledeg support from IRCICA (USR 3380 Univ. Lille - CNRS), from the ANR TOPWAVE (ANR-13-JS04-0004) project, from the "Fonds Europ\'{e}en de D\'{e}veloppement Economique R\'{e}gional", the Labex CEMPI (ANR-11-LABX-0007) and Equipex FLUX (ANR-11-EQPX-0017) through the "Programme Investissements d'Avenir".

%
%\bibliography{DarkBright}
%
%\appendix
%\section{derivation of Eq. \ref{eq:BSRR}}

\end{document}